\documentclass[twocolumn,letterpaper,amsmath,amssymb,floatfix,aps,superscriptaddress]{revtex4}
\usepackage{graphicx}
\usepackage{dcolumn}
\usepackage{bm}
\usepackage{epsfig,color}

\newcommand{\Av}[1]{{\bf #1}}

\def\ln{{\operatorname{ln}}}

\def\rmd{{\mathrm{d}}}

\def\rme{{\mathrm{e}}}
\def\lB{\ell_{\mathrm{B}}}

\begin{document}
\title{Attraction between Neutral Dielectrics Mediated by Multivalent Ions in an Asymmetric Ionic Fluid}

\author{Matej Kandu\v c}
\affiliation{Department of Physics, Technical University of Munich, D-85748 Garching, Germany}
\affiliation{Department of Physics, Free University Berlin, D-14195 Berlin, Germany}
\affiliation{Department of Theoretical Physics, J. Stefan Institute, SI-1000 Ljubljana, Slovenia}

\author{Ali Naji}
\affiliation{School of Physics, Institute for Research in Fundamental Sciences (IPM), Tehran 19395-5531, Iran}
\affiliation{Department of Applied Mathematics and Theoretical Physics, Centre for Mathematical Sciences, University of
Cambridge, Cambridge CB3 0WA, United Kingdom}

\author{Jan Forsman}
\affiliation{Department of Theoretical Chemistry, Lund University Chemical Center, P.O. Box 124, S-221 00 Lund, Sweden}

\author{Rudolf Podgornik}
\affiliation{Department of Theoretical Physics, J. Stefan Institute, SI-1000 Ljubljana, Slovenia}
\affiliation{Department of Physics, Faculty of Mathematics and Physics, University of Ljubljana, SI-1000 Ljubljana, Slovenia}
\affiliation{Department of Physics, University of Massachusetts, Amherst, MA 01003, USA}

\begin{abstract}
We study the interaction between two neutral plane-parallel dielectric bodies in the presence of a highly asymmetric ionic fluid, containing multivalent  as well as monovalent (salt) ions. Image charge interactions, due to dielectric discontinuities at the boundaries, as well as effects from ion confinement in the slit region between the surfaces are taken fully into account, leading to image-generated depletion attraction, ion correlation  attraction and steric-like repulsive interactions.  We investigate these effects by employing a combination of methods including explicit-ion and implicit-ion Monte-Carlo simulations, as well as an effective interaction potential analytical theory. The latter incorporates strong  ion-image charge correlations, which develop in the presence of high valency ions in the mixture. We show that the implicit-ion simulations and the proposed analytical theory can describe the explicit simulation results on a qualitative level, while excellent quantitative agreement can be obtained for sufficiently large monovalent salt concentrations. The resultant attractive interaction between the neutral surfaces is shown to be significant, as compared with the usual van der Waals interactions between semi-infinite dielectrics, and can thus play a significant role at the nano scale. 
\end{abstract}
\maketitle


\thispagestyle{plain}
\pagestyle{plain}

\pagenumbering{arabic}
\setlength\arraycolsep{2pt}

\section{Introduction} 

Apart from van der Waals (vdW) interactions, no universal, and certainly no electrostatic interactions are commonly assumed to exist between  uncharged (neutral) objects~\cite{French,Parsegian,Ninham-book}. Nevertheless, {\em some interactions} are every so often observed between neutral surfaces in the presence of bathing ionic (salt) solutions~\cite{Hydration} and Wernersson \& Kjellander studied theoretically the case of a salt solution with mono- and/or divalent ions confined in a slit between neutral dielectrics by employing the anisotropic hypernetted chain method~\cite{Kjellander1,Kjellander0}. Electrostatic interactions, mediated by mobile monovalent ionic solutions, are known to be described very well in terms of mean-field theories such as the Poisson-Boltzmann (PB) equation~\cite{leshouches}. For the particular case of symmetric uncharged plane-parallel dielectric bodies, which will be considered in this paper, the standard PB treatment cannot account for inter-surface forces as it yields only a trivial uniform ion distribution in the slit between the inner surfaces of the slabs, and thus a nil result for the inter-surface force. However, in the absence of surface charges, other electrostatic effects involving mobile ions, not included in the PB approach, increase in importance, especially  when multivalent ions are present in the slit. These include fluctuation/correlation and, in particular, dielectric image effects, which are missing within the standard  PB description in planar geometry~\cite{Andelman}. 
In fact,  correlation effects are expected to be significant at uncharged interfaces, since for each individual multivalent ion,  the screening cloud and its dielectric images are pinned to the mobile ion itself rather than any fixed external surface charges. Such  ``ion-image" correlations constitute the core insight of our subsequent analysis  for neutral surfaces, suggesting that a proper theoretical understanding of interactions between neutral surfaces may thus follow from the fact that ions and their dielectric images are {\em strongly correlated}. 

Such correlation effects are expected to be important, especially in the presence of highly asymmetric salt mixtures containing a (usually) low concentration of an ionic species with a high valency.  This latter situation is in fact quite relevant experimentally~\cite{rau-1,rau-2}. We thus focus on such asymmetric salt mixtures confined between two uncharged plane-parallel dielectric half-spaces, and extend the existing analysis~\cite{Kjellander1} to the case where the asymmetric ionic mixture contains at least one multivalent ion component of high valency. The case of symmetric ionic fluids and/or weakly charged ionic mixtures at neutral interfaces, differing from the focus of the present work, has been approached on various levels of approximations \cite{Ninham-book,Kjellander0,lue,Netz0,Netz1,Netz2} starting from the seminal theoretical work of Onsager and Samaras \cite{OS}.

We base our approach on a combination of methods, i.e., explicit-ion, implicit-ion and Yukawa Monte-Carlo (MC) simulations as well as an effective ``dressed multivalent ion" theory, which accounts for the salt screening effects as well as the dielectric polarizability of the neutral dielectrics. We show that this latter approach provides a simple and useful analytical framework for the present and previous~\cite{Kjellander1} results. It is similar to a  ``dressed counterion" model discussed in the context of charged surfaces in contact with a bathing salt solution containing multivalent  ``counterions"~\cite{kanduc_dressed1,kanduc_dressed2}. There it was shown that for high-valency counterions, the degrees of freedom associated with the monovalent ions can be integrated out and an effective single-particle theory can be obtained based on a virial expansion in terms of the counterions fugacity, which agrees very well with MC simulations within its regime of applicability. This analysis is therefore also akin to the limiting single-particle description, known as the strong coupling (SC) theory, which was introduced originally for counterion-only systems~\cite{netz,moreira2000,hoda,physicaA,revali,kanduc2007,jhoPRL,olli}. The SC theory incorporates, to the leading order, strong electrostatic correlations, which develop between counterions and the surface charges in the limit of large counterion valency. For uncharged surfaces, such surface charge-counterion correlations are absent but, nevertheless, the presence of multivalent ions can make the collective mean-field description obsolete, provided that the surface dielectric (``image charge") effects are properly accounted for. As we will show, the single-particle SC framework can be reformulated and  generalized in order to capture the ion-image correlation effects and thereby, the interaction mediated by an asymmetric salt mixture between two uncharged dielectric bodies, even on a quantitative level. 

These interactions are shown to be quite short-range (up to separation distances of just a few nano-meters) 
and are of electrostatic nature even for neutral bodies. They combine depletion as well as dielectric image effects in a novel kind of interaction  that can play a significant role in shaping the behavior of neutral soft matter systems--most prominently at the nano scale. These interactions should be therefore considered in parallel with 
the hydrophobic interactions~\cite{hydrophobic} as well as vdW interactions~\cite{Parsegian} deemed to play an important role in the interaction of neutral surfaces.

We base our analysis on explicit-ion MC simulations of non-specific ion interactions 
and while we take dielectric image interactions properly into account, we disregard the vdW (dispersion) interactions of the ions with the dielectric interfaces, an approximation that is justified for  high-valency ions~\cite{Kjellander1}. The purpose and emphasis of our endeavor is thus twofold:  to show that, for highly asymmetric ionic mixtures, a coarse-grained MC approach  with {\em implicit} monovalent ions can be properly applied to describe interactions between neutral surfaces, and secondly to provide a simple analytical theory, which could explain the main features of these simulations on a qualitative level that becomes quantitative at sufficiently high monovalent salt concentration.

\begin{figure}[t!]\begin{center}
	\begin{minipage}[b]{0.5\textwidth}\begin{center}
		\includegraphics[width=\textwidth]{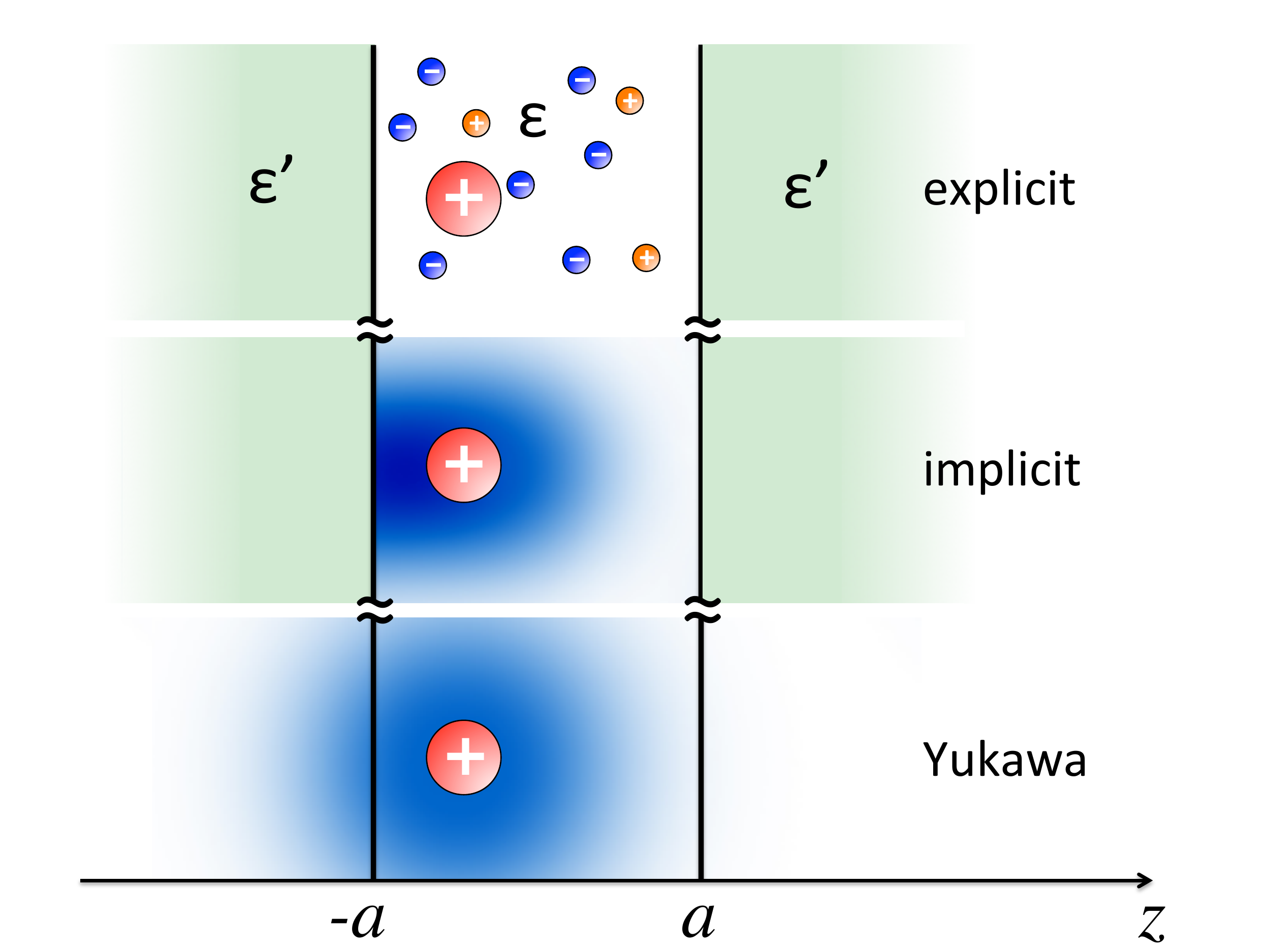}
	\end{center}\end{minipage} 
\caption{(Color online) Schematic depiction of two neutral dielectric half-spaces at separation $2a$. The slit region is in equilibrium with a bulk reservoir containing a monovalent 1:1 salt as well as an asymmetric multivalent $q$:1 salt. Explicit-ion, implicit-ion and Yukawa levels of description are indicated schematically (top to bottom)--see the text for details.}
\label{fig:geom00}
\end{center}\end{figure}

\section{Model and methods} 
\label{sec:model}

Let us consider two neutral plane-parallel dielectric half-spaces of dielectric constant  $\varepsilon'$ at a separation distance of $2a$, as shown schematically in Fig~\ref{fig:geom00}. The intervening slit is assumed to be 
filled with an aqueous solution of dielectric constant  $\varepsilon$, containing an ionic mixture comprised of a 1:1 monovalent salt solution of bulk concentration $n_0$ and an asymmetric $q$:1 multivalent salt solution of bulk concentration $c_0$, where $q$ stands for the charge valency of multivalent ions. The interfacial dielectric contrast can be quantified by a {\em dielectric mismatch parameter}, $\Delta=({\varepsilon-\varepsilon'})/({\varepsilon+\varepsilon'})$. 
In order to  investigate the inter-surface forces induced between neutral dielectrics within the model described above, we shall employ MC simulations on two different levels of description as well as an analytical theory. Specifically, we use {\em explicit-ion simulations} (referred to as  ``explicit" simulations) where all ions in the ionic mixture, including those
of the 1:1 and $q$:1 salt, are simulated explicitly (see Fig.~\ref{fig:geom00}, top). In this case, the closest-approach distances between the ions are taken as $d_{+1,-1}=0.35$~nm between monovalent cations and anions, and $d_{q,\pm1}=0.45$~nm between multivalent ions and monovalent ions. 
These values are  small enough so that the excluded-volume effects are negligible, yet large enough to prevent Bjerrum pairing of ions in the solution.  The dielectric polarization of the half-spaces with planar boundaries 
is accounted for by using repeated image reflections. The image reflection sum has to be truncated for computational reasons, and we found satisfactory accuracy with only three explicit reflections  across each surface. All MC simulations were performed using a multi-component grand-canonical ensemble, ensuring equilibrium with a bulk solution of the prescribed composition.  

In the second approach, we use  {\em implicit-ion simulations} (referred to as  ``implicit" simulations)  
in which only multivalent ions ($q>1$) are explicitly simulated, whereas the effects of monovalent ions are included implicitly via a Debye-H\"uckel-like ionic cloud. In the implicit model, apart from the dielectric polarization of the half-spaces,  we account for the fact that the screening ionic cloud around multivalent ions is not spherically symmetric but distorted near the (hard) boundaries (see Fig.~\ref{fig:geom00}, middle).  This is accomplished by using the modified non-homogeneous Debye-H\"uckel (DH) interaction kernel (Green's function) $u_\textrm{DH}(\Av r,\Av r')$ in order to compute the electrostatic interaction energy between each two (explicit) multivalent ions of electric charge $qe_0$ located at positions $\Av r$ and $\Av r'$ in the slit, i.e., $(qe_0)^2u_\textrm{DH}(\Av r,\Av r')$. In the Fourier-Bessel representation, $u_\textrm{DH}(\Av r,\Av r')$  can be written as
\begin{equation}
u_\textrm{DH}(Q;z,z')=u_\textrm{DH}^{0}(Q;z,z')+u_\textrm{DH}^\textrm{im}(Q;z,z'), 
\label{eq:u}
\end{equation}
where $Q$ is the norm of the Fourier wave-vector corresponding to a two-dimensional Fourier-Bessel transform with respect to the lateral directions $\boldsymbol\rho=(x, y)$ perpendicular to $z$-axis.  The first term above is the isotropic (bulk)  ``Yukawa" potential and the second term, usually ascribed to ``image" charges, accounts for the presence of solid planar dielectric walls. It consistently incorporates the dielectric polarization of the half-spaces as well as  the effects due to confinement of the salt ions to the slit only, preventing the spillage of their DH cloud into the bounding dielectric half-spaces. 
This last effect is in fact an essential component of the interaction between the surfaces. 
Neglecting  this kind of DH cloud distortions can lead to qualitatively wrong predictions, as we shall demonstrate. 
In explicit form, the Fourier-Bessel transforms are given by 
\begin{eqnarray}
u_\textrm{DH}^\textrm{0}(Q;z,z')\!&=&\!\frac{1}{2\varepsilon\varepsilon_0 k}\,\rme^{-k\vert z-z'\vert},
\label{eq:u0}\\
u_\textrm{DH}^\textrm{im}(Q;z,z')\!&=&\!\frac{\Gamma(Q)\cosh k(z-z')+\cosh k(z+z')}{\varepsilon\varepsilon_0 k\,[\Gamma^{-1}(Q)-\Gamma(Q)]},\hspace{3ex}
\label{eq:uim}
\end{eqnarray}
where
\begin{eqnarray}
\Gamma(Q)&=&\frac{(1+\Delta)k-(1-\Delta)Q}{(1+\Delta)k+(1-\Delta)Q}\,\rme^{-2ka}, 
\label{eq:Gamma}
\end{eqnarray}
and  $k=\sqrt{\kappa^2+Q^2}$ with $\kappa$ being the Debye screening parameter defined via 
$\kappa^2=4\pi\lB n_b$.  Here $\lB = e_0^2/(4\pi \varepsilon\varepsilon_0 k_{\mathrm{B}}T)$ is the Bjerrum length at ambient temperature $T$ and 
\begin{equation}
n_b= 2n_0+ qc_0
\end{equation} 
is the {\em total} density of all monovalent ions in the bulk. Note that when multivalent $q$:1 salt is introduced in the system,  additional monovalent ions with concentration $q c_0$ are simultaneously released into the bulk and thus they also contribute to the screening  parameter~\cite{kanduc_dressed2,chan}.  

The back-transform of the interaction kernel to real space follows from
\begin{equation}
u_\textrm{DH}(\Av r, \Av r')=\frac 1{2\pi}\int_0^{\infty}Q\, \rmd Q \,J_0(Q|\boldsymbol\rho-\boldsymbol\rho'|)\,u_\textrm{DH}(Q; z,z'),
\end{equation}
where $J_0(x)$ is the zeroth-order Bessel function of the first kind. In the limit of high dielectric mismatch, $\Delta\to 1$, the interaction kernel can be  simplified significantly, yielding an analytical form in real space as
\begin{eqnarray}
u_\textrm{DH}(\Av r, \Av r')&=&\frac 1{4\pi\varepsilon\varepsilon_0}\biggl[\,\sum_{n \textrm{ even}}\frac{\rme^{-\kappa|\Av r'-\Av r-2na\,\Av e_z|}}{|\Av r'-\Av r-2na\,\Av e_z|}\nonumber\\
&&\quad+\sum_{n \textrm{ odd}}\frac{\rme^{-\kappa|\Av r'-\Av r-2(na-z')\Av e_z|}}{|\Av r'-\Av r-2(na-z')\Av e_z|}\,\biggr], 
\end{eqnarray}
where $\Av e_z = (0,0,1)$ is the unit vector in the $+z$ direction and the integer index $n$ runs through both positive and negative values. The term with $n=0$ stands for the direct DH potential, $u_\textrm{DH}^\textrm{0}(\Av r, \Av r')$, of a test charge without any dielectric effects. The above expression thus reflects a simple picture within which the total potential between two parallel dielectric half-spaces can be represented as a sum of an infinite number of screened Coulomb image charges. This form of image summation significantly reduces also the computational costs, since summing up to, e.g.,  $n=5$, already gives a very well-converged result for the cases we  consider in this work. We have thus used the above summation series in the implicit MC simulations, and focus on the dielectric jump value $\Delta=0.95$, corresponding to the water-hydrocarbon interface (assuming $\varepsilon=80$ and $\varepsilon'=2$). The overall Hamiltonian in the implicit MC approach then reads,
\begin{equation}
H =\sum_{i>j}(qe_0)^2 u_\textrm{DH}(\Av r_i, \Av r_j)+\frac 12\sum_{i}(qe_0)^2 u_\textrm{DH}^\textrm{im}(\Av r_i, \Av r_i), 
\end{equation}
combining the screened interaction with distorted DH cloud between all multivalent ion pairs and their images (first term) as well as their self-image interaction energy (second term).

The implicit MC simulations were performed in the grand-canonical ensemble. The simulated lateral system size was chosen such that the slit contained between around 50 to 200 multivalent ions.

In a third simulation approach (referred to as  ``Yukawa" simulations), we use the same implicit model as above but neglect the distortion of the ionic clouds near the half-spaces as well as their polarizability (see Fig.~\ref{fig:geom00}, bottom). We thus use the fully symmetric screened Coulomb or Yukawa kernel (\ref{eq:u0}), which in real space reads
\begin{equation}
u_\textrm{DH}^\textrm{0}(\Av r, \Av r')=\frac{\rme^{-\kappa |\Av r-\Av r'|}}{4\pi\varepsilon\varepsilon_0 |\Av r-\Av r'|},
\end{equation}
 in order to compute the electrostatic interactions between multivalent ions.

\begin{figure*}[t]\begin{center}
\begin{minipage}[b]{0.32\textwidth}\begin{center}
\includegraphics[width=\textwidth]{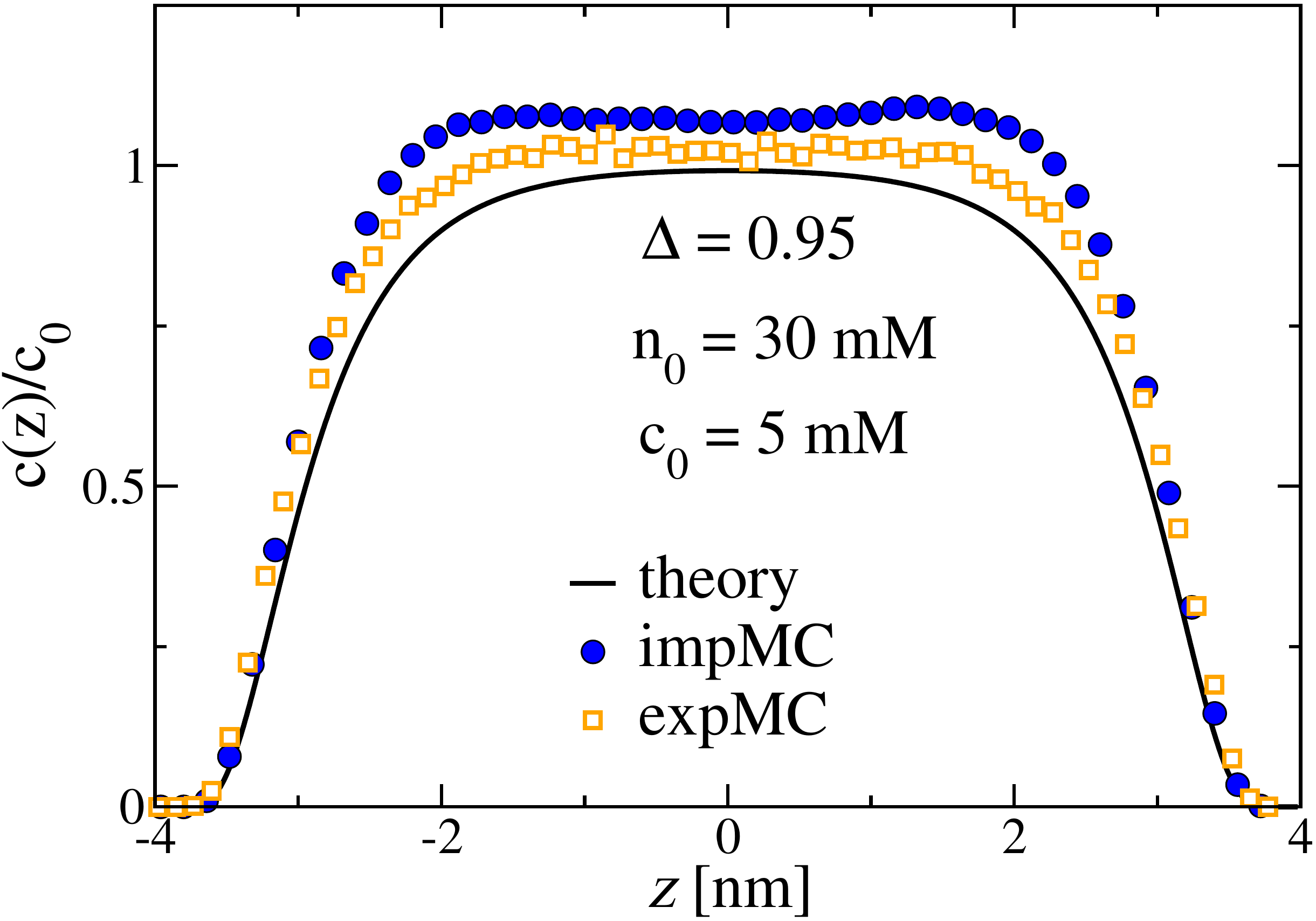} (a)
\end{center}\end{minipage}
\begin{minipage}[b]{0.32\textwidth}\begin{center}
\includegraphics[width=\textwidth]{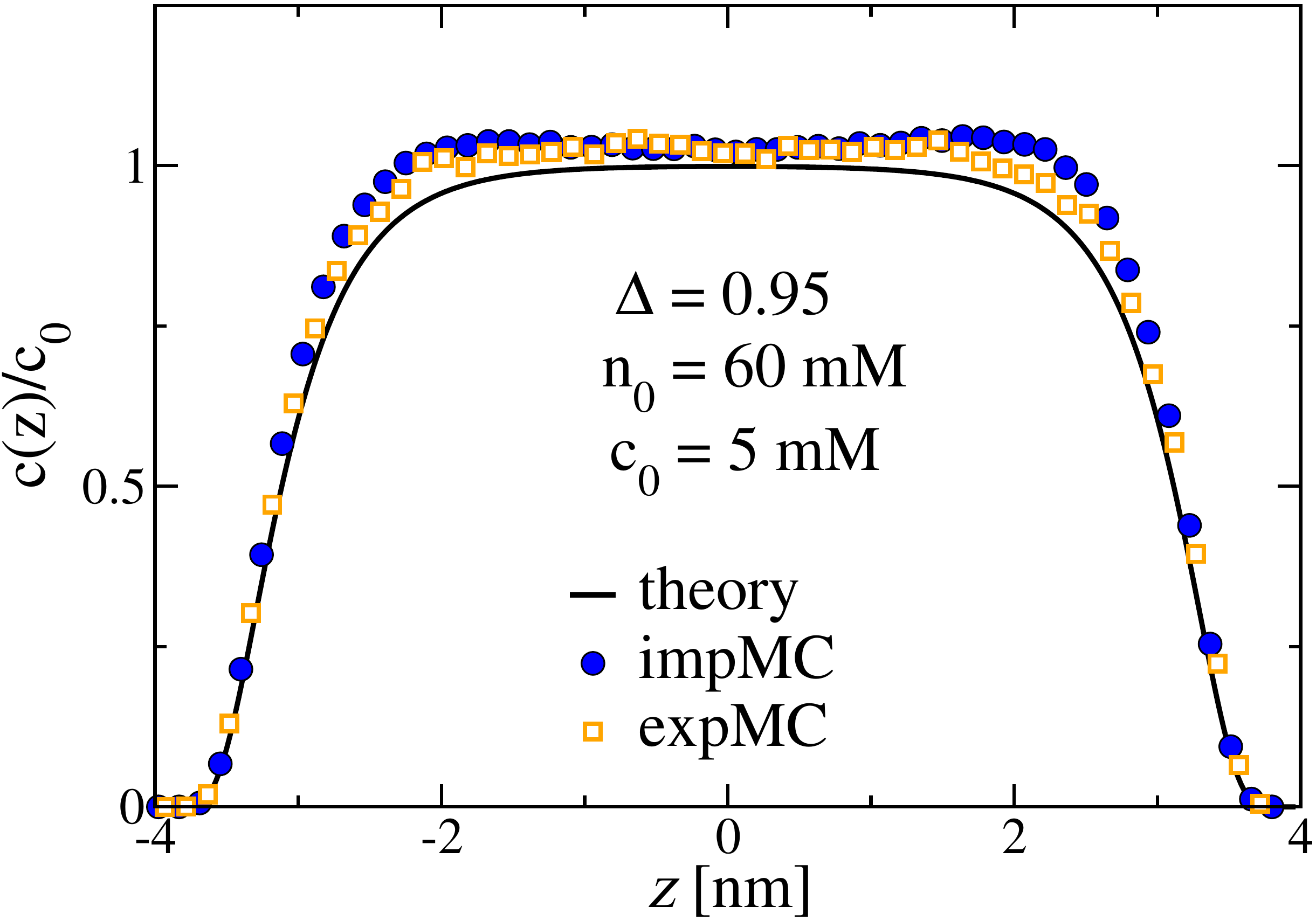} (b)
\end{center}\end{minipage}
\begin{minipage}[b]{0.32\textwidth}\begin{center}
\includegraphics[width=\textwidth]{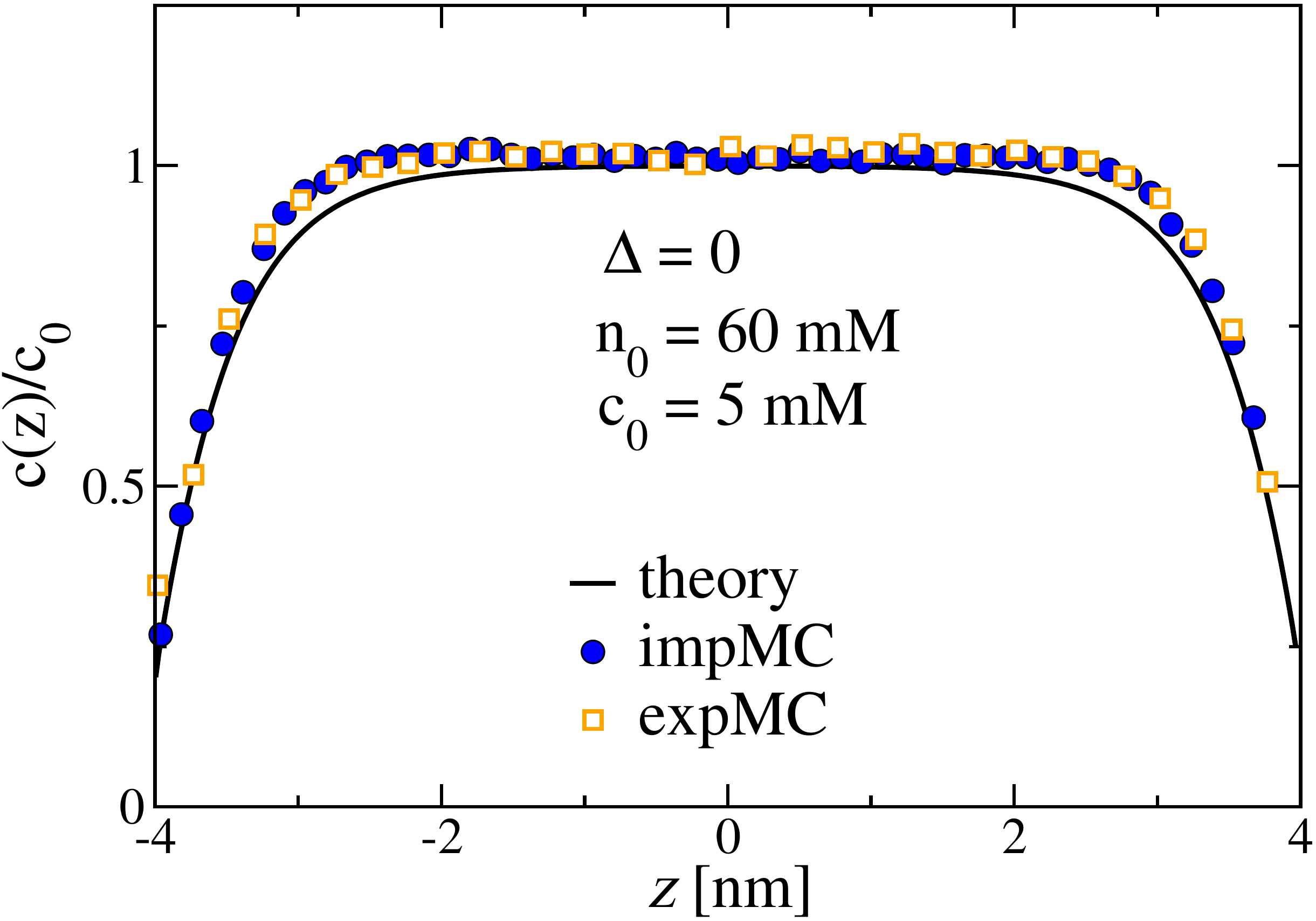} (c)
\end{center}\end{minipage}
\caption{(Color online) Rescaled density profiles of multivalent ions ($q=4$) in the slit between two neutral half-spaces at separation $2a=8$\,nm.  The bulk concentration of multivalent 4:1 salt is taken as $c_0=5$~mM and the bulk concentration of the additional monovalent salt is $n_0=30$~mM  in pane (a) and  $n_0=60$~mM in panes (b) and (c). We show the results from our explicit MC simulations (orange open squares), implicit MC simulations (blue filled circles) and the proposed analytical theory (solid curves). In (a) and (b)  the dielectric mismatch at the  boundaries is $\Delta=0.95$, whereas in (c) there are no dielectric discontinuities, $\Delta=0$. 
The dielectric surfaces are assumed to be impermeable to all ions.}
\label{fig:dens}
\end{center}\end{figure*}

\section{Results}

\subsection{Density profile}

In Fig.~\ref{fig:dens}, we show the density profiles of multivalent ions with valency $q=4$ within the slit, as obtained from our MC simulations  for the case of a 4:1 and 1:1 salt mixture. In all cases  the multivalent salt is assumed to have a bulk concentration of $c_0=5$~mM. In Fig.~\ref{fig:dens}a, we have an additional 1:1 salt of bulk concentration $n_0=30$~mM as well as a dielectric jump parameter of $\Delta=0.95$.  As seen, both explicit and implicit MC simulations show a pronounced depletion of multivalent ions from the proximity  of the two neutral walls, where the density tends to zero. The explicit and implicit data nearly agree, particularly near the walls, with somewhat larger deviations in the middle of the slit. A similar surface depletion effect can be seen when the additional monovalent salt is increased but the agreement becomes progressively better; in Fig.~\ref{fig:dens}b,  we show the results for $n_0=60$~mM, which exhibits a very good agreement between the  explicit and implicit MC simulations. This confirms the idea that the degrees of freedom associated with monovalent ions enter only in an effective Debye-H\"uckel way as formulated in Eqs. (\ref{eq:u})-(\ref{eq:Gamma}). Therefore the depletion of multivalent ions  can indeed be ascribed both to dielectric images due to dielectric polarization of the half-spaces as well as to the fact that the exclusion of salt ions from the impenetrable surfaces leads to a non-trivial distortion of ionic clouds in the vicinity of the half-spaces (Fig.~\ref{fig:geom00}, middle), effectively engendering ``salt-induced image charges" which repel the ions from the walls.  Alternatively, one could say that ions in the vicinity of the interfaces have fewer neighbors and thus all the ions, i.e., both monovalent (not shown) \cite{mono}  and multivalent ions, are depleted from the proximity of the surfaces.  Such  ``salt-image effects" are present even if there is no dielectric discontinuity at the boundaries ($\Delta=0$) as seen in Fig.~\ref{fig:dens}c, and lead to a significant depletion near the surfaces, albeit less pronounced than in the presence  of standard dielectric image effects ($\Delta>0$, Figs.~\ref{fig:dens}a and b).  Other remarkable differences with the latter case are that  multivalent ions can have a finite contact density at the boundaries when $\Delta=0$  and also exhibit 
an even better agreement between explicit and implicit MC data (Fig.~\ref{fig:dens}c).

The above results clearly go beyond the standard PB description which is expected to give only a trivial uniform ion distribution between neutral surfaces in the planar geometry.  In order to gain any insight into the behavior of this system one needs to conceptualize outside of the PB box. We thus proceed by proposing a simple analytical theory, akin to the  concept of  ``dressed counterions", which was developed previously in the context of a multivalent salt solution confined between two charged surfaces without a dielectric jump~\cite{kanduc_dressed1} or next to one charged dielectric interface~\cite{kanduc_dressed2}, where it was exhaustively and successfully tested against MC simulations. In the present context with neutral surfaces, where one can not speak of  ``counter ions" proper, we refer to this approximation as the  ``dressed multivalent ion" approach. It follows from the same principles as our implicit MC simulations, i.e., by integrating out the degrees of freedom associated with all monovalent ions treated on a linear DH level, and hence by obtaining an effective theory where multivalent ions are still described explicitly but the role of monovalent ions enters only implicitly~\cite{kanduc_dressed1}. This effective theory  can be justified only for highly asymmetric ionic mixtures  with large $q$, as is of primary interest here, and can be obtained systematically from a field-theoretical formalism, which we shall not discuss further in this paper (see, e.g., Ref.~\cite{kanduc_dressed1} for details). Note that the system of dressed multivalent ions still presents a full many-body problem and as such can be treated exactly only via simulations (which, in the present context, would  correspond precisely to our implicit MC simulations that incorporate the proper implicit interaction kernels). An approximate analytical theory then follows from a virial expansion, which for neutral systems is expected to work at sufficiently small concentration of multivalent ions, as is often  the case in experimental systems~\cite{rau-1,rau-2}.  For charged surfaces,  such a virial expansion can be shown to be related to a strong-coupling expansion of multivalent counterions~\cite{netz}, where electrostatic correlations between counterions and the opposite surface charges become dominant on the leading order. In the present case, such a strong-coupling theory for multivalent ions can not be developed as the surfaces are neutral and the electrostatic coupling parameter {\em \' a la Netz} is non-existent. Nevertheless, the virial expansion can capture the salient features of the problem as it incorporates the ion-image correlation, which in this context becomes the dominant mechanism at work for multivalent ions at low concentrations. 

On the leading order one thus finds that the only driving force for redistribution of 
multivalent ions in the slit $|z|<a$ is their self-energy, leading to the density profile 
\begin{equation}
c(z)=c_0\,\rme^{-q^2w(a,z)}, 
\label{eq:n_multi}
\end{equation}
where the self-energy (per unit $k_{\mathrm{B}}T=\beta^{-1}$)  reads
\begin{equation}
w(a, z) \equiv \beta e_0^2 u_\textrm{DH}(\Av r,\Av r) = \lB\!\!\int_0^\infty\frac{\Gamma(Q)+\cosh 2kz}{\Gamma^{-1}(Q)-\Gamma(Q)}\,\frac{Q\,\rmd Q}{k}.
\end{equation}
The above analytical result is shown in Fig.~\ref{fig:dens} (solid curves) and appears to capture the main trends of the MC data. A close agreement between this theory and the simulations is obtained only for large enough  monovalent salt concentration (Fig.~\ref{fig:dens}b) or when the dielectric jump is small (Fig.~\ref{fig:dens}c). In the case $\Delta\to 1$, the above self-energy expression simplifies into
\begin{eqnarray}
w(a,z)&=&-\frac{\lB}{4a}\,\ln(1-\rme^{-4\kappa a})\\
&&+\frac 12\lB\sum_{m\textrm{ odd}}\frac{ma\,\cosh 2\kappa z+z\sinh 2\kappa z}{m^2a^2-z^2}\,\rme^{-2m\kappa a}.\nonumber
\end{eqnarray}
where $m=1,3,5,\ldots$, and thus allows for an analogy with the Kelvin image charges.

It should be noted that the above analytical theory predicts multivalent ion densities in the slit that are slightly smaller than the bulk value $c_0$ (see Fig.~\ref{fig:dens}).
In contrast, the simulations do reveal an additional buildup of the concentration of multivalent ions in the slit 
that are forced in due to their mutual repulsive interactions in the bulk, an effect
which is absent in the theory.

\begin{figure}[t]\begin{center}
\begin{minipage}[b]{0.4\textwidth}\begin{center}
	\includegraphics[width=\textwidth]{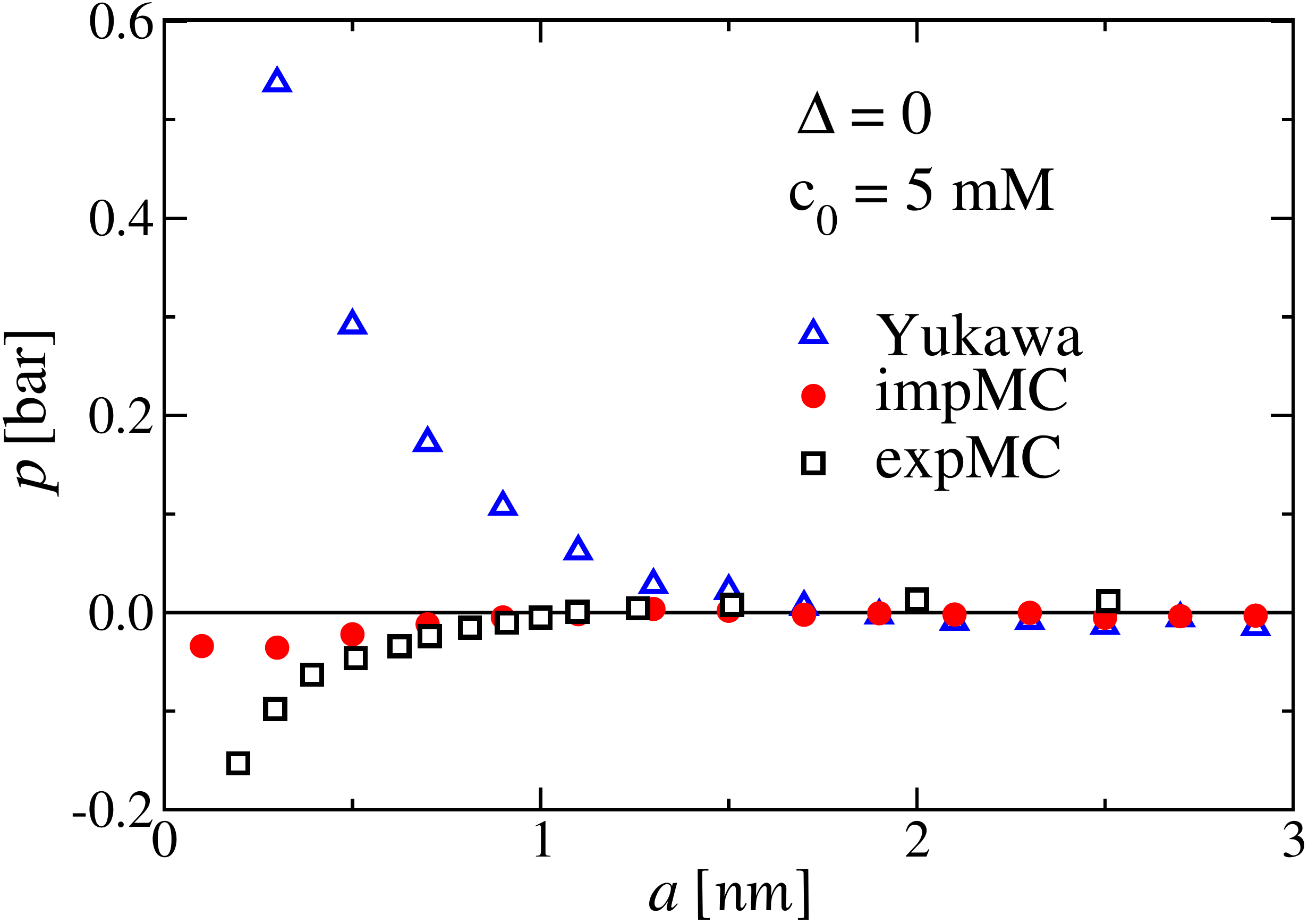} 
\end{center}\end{minipage} 
\caption{(Color online) MC results for the interaction pressure mediated between two neutral dielectric surfaces as a function of their separation in the presence of a multivalent 4:1 salt mixture of bulk concentration $c_0=5$~mM and with no additional monovalent salt ($n_0=0$) and no dielectric discontinuities ($\Delta=0$). We compare
the naive Yukawa approach with the implicit and explicit MC simulations. 
}
\label{fig:Yukawa}
\end{center}\end{figure}

\subsection{Interaction pressure}

We now turn our attention to the interaction (disjoining) pressure acting on each of the neutral surfaces due to the presence of the salt mixture in the slit.  This pressure is composed of two separate additive contributions: one stemming from the monovalent ions as constituents of the 1:1 salt as well as of the  $q$:1 salt, and  the other originating from multivalent ions. The latter one can be calculated analytically from the single-particle grand canonical potential within the  ``dressed multivalent ion" approximation. The former is trickier and can only be evaluated approximately by starting from an exact pressure equation for a simple salt system and then by remorphing it in a self-consistent way, using the same line of thought as in the case of multivalent ions. This is reasonable as the non-homogeneous interaction kernel, Eq. (\ref{eq:u}), applies equally to interaction between any two charges, as can be clearly deduced from detailed MC simulations \cite{mono}.

\begin{figure*}[t]\begin{center}
\begin{minipage}[b]{0.31\textwidth}\begin{center}
\includegraphics[width=\textwidth]{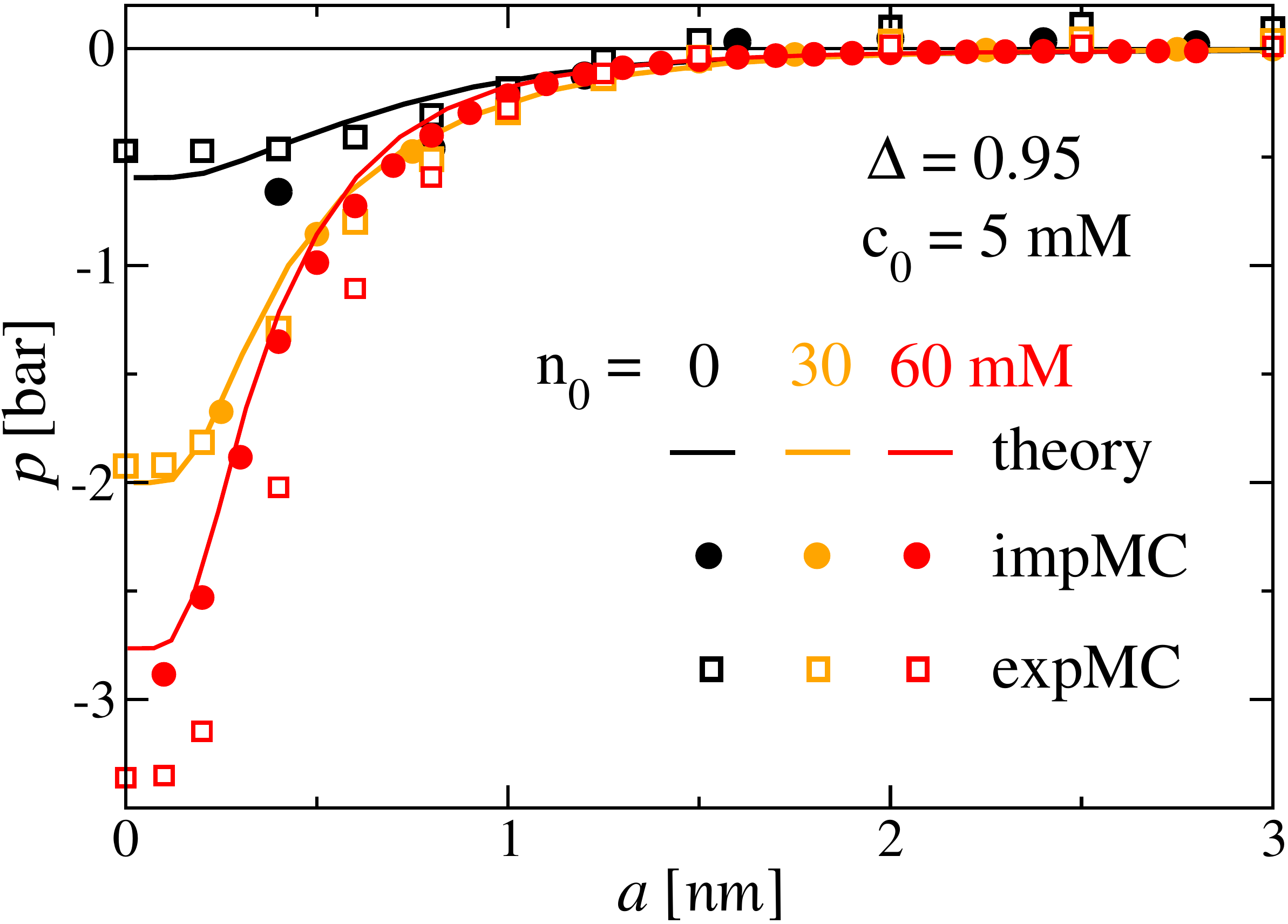} (a)
\end{center}\end{minipage}
\begin{minipage}[b]{0.33\textwidth}\begin{center}
\includegraphics[width=\textwidth]{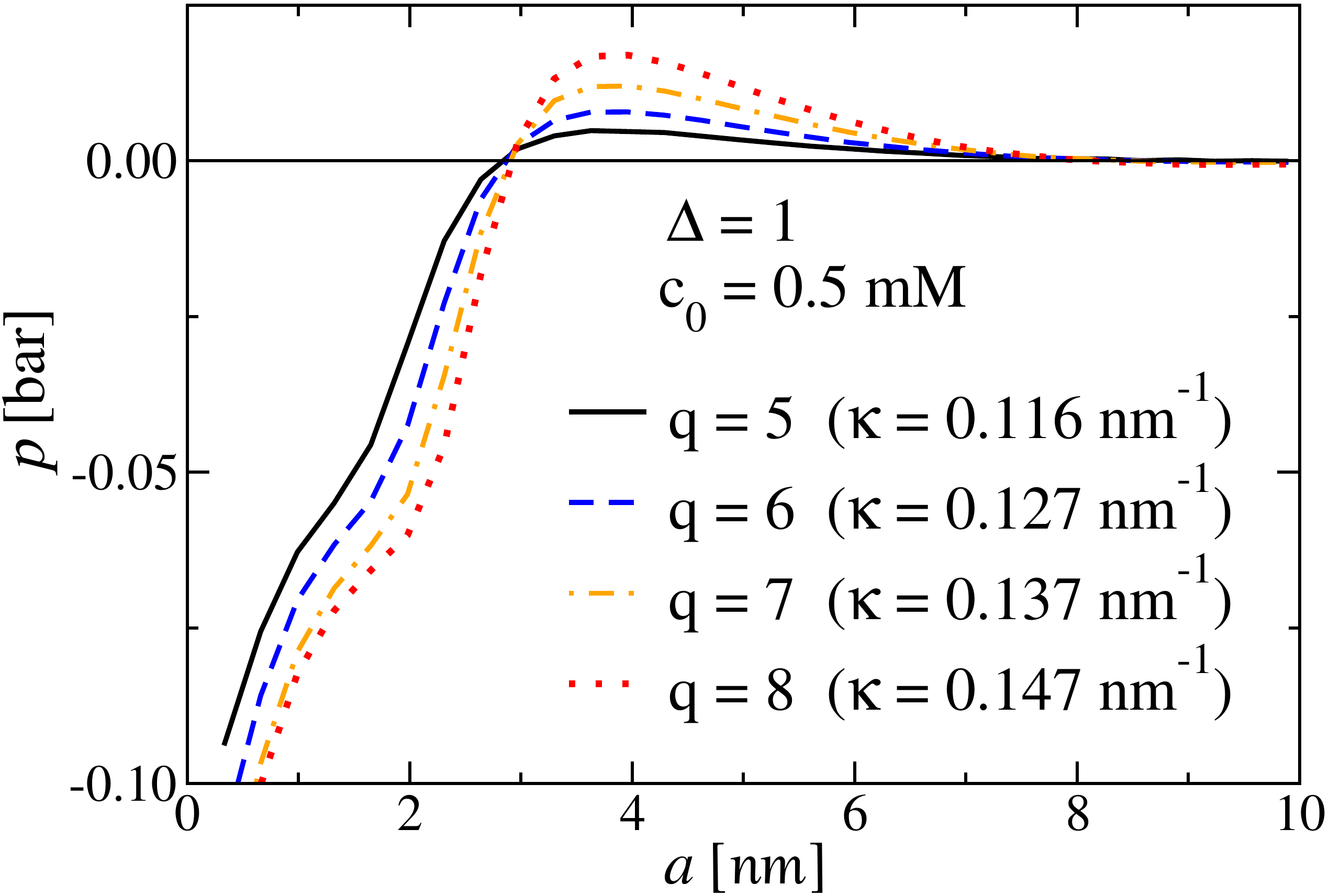} (b)
\end{center}\end{minipage}
\begin{minipage}[b]{0.33\textwidth}\begin{center}
	\includegraphics[width=\textwidth]{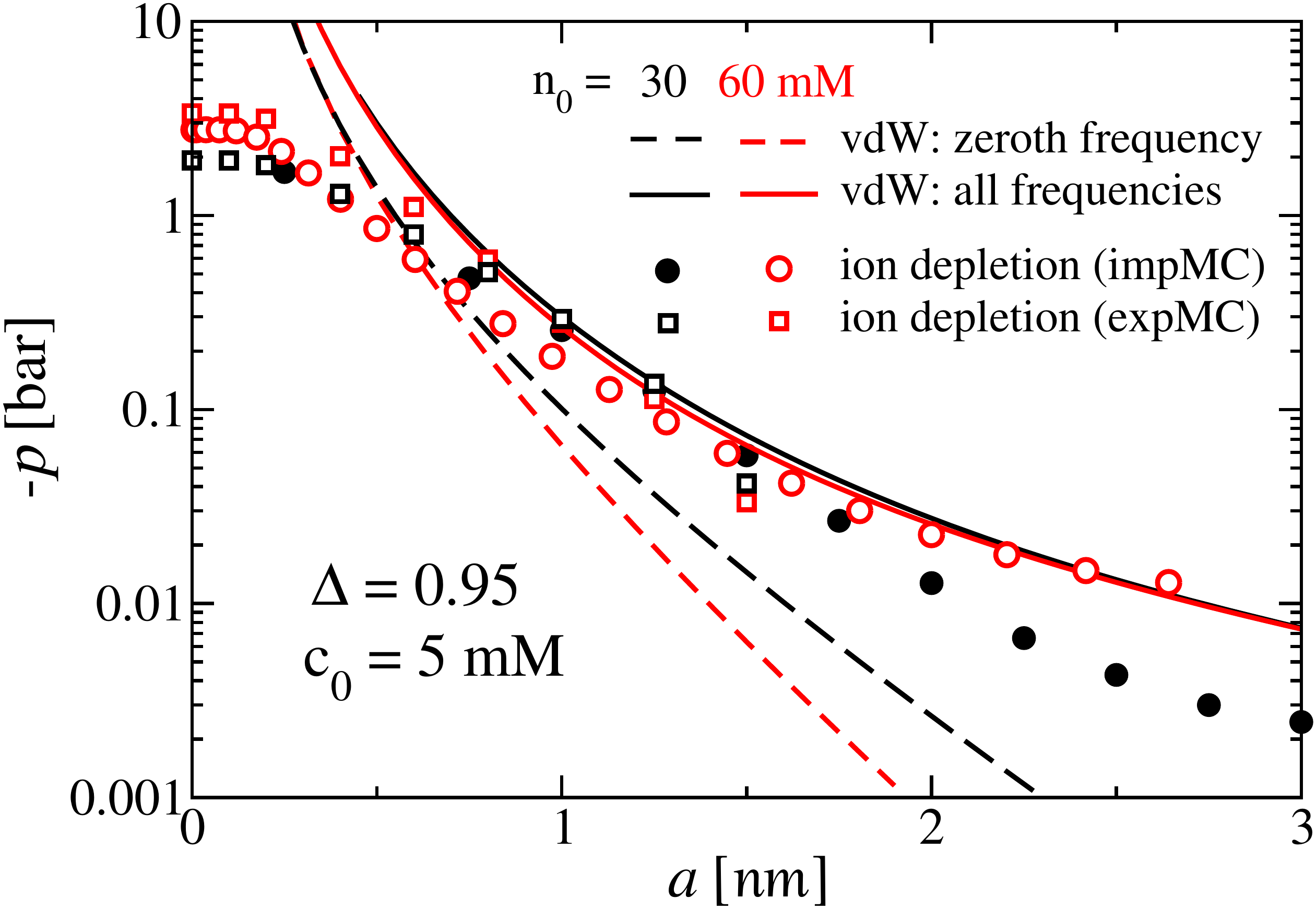} (c)
\end{center}\end{minipage} 
\caption{(Color online) Pressure acting between the surfaces of two neutral dielectric half-spaces as a function of their separation distance for various amounts of salt concentrations. (a) We compare the results from the explicit MC simulations (open squares), implicit MC simulations (filled circles), and the proposed analytical theory (solid curves) for a mixture of multivalent 4:1 salt with bulk concentration $c_0=5$\,mM and different additional 1:1 salt concentrations of $n_0=0$ (black), 30\,mM (orange) and 60\,mM (red).  In (b) we show the implicit MC results for $c_0=0.5$~mM and various multivalent ion valency $q$ as shown on the graph. In (c) we compare the attractive pressures obtained from explicit (squares) and implicit (circles) MC simulations with the contribution from the zero-frequency vdW (dashed curves) and full vdW (solid curves) interactions for the cases with  $c_0=5$\,mM and $n_0=30$\,mM (black) and 60\,mM (red). All our results are on top of the underlying vdW interactions.}
\label{fig:pressure}
\end{center}\end{figure*}

In the case of implicit MC simulations as well as the proposed analytical approach, the  contribution of monovalent ions to the slit pressure, $p_0$, can be derived from an exact equation formulated by Lovett and Baus~\cite{lovett}, which is closely related to the so-called contact-value theorem in the plane-parallel geometry~\cite{leshouches,Andelman}, and can be shown to have the following approximate linearized (DH) form 
\begin{eqnarray}
\beta p_0(a)&\simeq &n(0)-\frac{(\beta e_0^2)^2}{2\pi}\int_{-a}^0\int_0^a \rmd z_1\, \rmd z_2\,
n(z_1)n(z_2)\times \nonumber\\
&&\hspace{-1.5cm}\times\int_0^\infty Q\,\rmd Q\, u_\textrm{DH}(Q;z_1,z_2)\, 
\frac{\partial}{\partial z_1}u_\textrm{DH}(Q;z_1,z_2)\big|_{\kappa=0}.
\label{eq:contact2}
\end{eqnarray}
 It is composed of a van't Hoff term (first term) due to the ion osmotic pressure and a correlation term (second term) due to the interaction between the ions. 
 It should be noted that monovalent ions do also 
 exhibit depletion (even though much weaker than the multivalent ions) from the vicinity of the walls, and thus a non-uniform density profile in the slit.  This effect is implicitly embedded in the modified DH interaction kernel as noted 
 above (see also Ref.  \cite{Netz2}), and has been verified within our explicit MC simulations \cite{mono}. 
We find that the {\em total} density profile  of monovalent ions can be described very well by their corresponding self-energy as
\begin{equation}
n(z)=n_b\,\rme^{-w(a,z)},
\end{equation}
which is to be used in expression (\ref{eq:contact2}) for the depletion pressure due to monovalent ions.  

In order to evaluate the slit pressure due to multivalent ions analytically, we now use the virial
scheme, which to the leading order is given by a single-particle grand canonical potential $\beta\Phi^{(1)}$ and can be calculated explicitly~\cite{kanduc_dressed1,kanduc_dressed2}. 
The slit pressure can be deduced by differentiation with respect to 
the volume of the slit, $p_\textrm m=-(\partial\Phi^{(1)}/\partial V)_{\beta, \lambda}$, 
where $ \lambda$ is the fugacity of the multivalent ions. We thus find
\begin{equation}
\beta p_{\textrm{m}}(a)=\frac{c_0}{2} \frac{\partial}{\partial a}\int_{-a}^a\rmd z\, \rme^{-q^2w(a,z)}.
\label{eq:pcSC}
\end{equation}
This contribution should be added to the one from the monovalent  ions in order to obtain the total slit pressure. 
The  {\em interaction} (disjoining) pressure acting on the bounding surfaces follows by subtracting the bulk pressure, $p_\textrm{bulk}$, as
\begin{equation}
p(a)=p_0(a)+p_\textrm m(a) - p_\textrm{bulk},
\label{eq:ptot}
\end{equation}
where  $p_\textrm{bulk} = p_0(a\rightarrow \infty)+p_\textrm m(a\rightarrow \infty)$. 

In order to proceed with the simulation results, we first consider the case with no dielectric mismatch across the bounding surface, i.e., $\Delta=0$. In Fig.~\ref{fig:Yukawa},  we show our MC simulation results from the explicit, implicit as well as Yukawa approaches (see Section \ref{sec:model}) for the case where the slit contains only a multivalent 4:1 salt mixture of bulk concentration $c_0=5$~mM and no additional monovalent salt ($n_0=0$).  As seen, the explicit and implicit simulations predict an attractive force at small separations which we shall  elaborate on further below. It is interesting to note however that the naive Yukawa  treatment of the multivalent salt without incorporating the image kernel, Eq.~(\ref{eq:uim}), leads to qualitatively incorrect results as it gives a repulsive interaction pressure.  The repulsion in this latter case appears to be purely osmotic in nature as surfaces are uncharged. In the case of implicit or explicit simulations, the interactions due to dielectric and ``salt-induced" image charges lead to depletion of ions from the slit which in turn engenders a depletion attraction between the surfaces. This depletion is thus generated by a confined ionic cloud in contrast to standard depletion interactions, which are steric in nature~\cite{lekkerkerker}. Both types of depletion effects are rather similar except that the image-depletion interaction leads to softer dependence of the interaction pressure on separation when compared to the standard hard-sphere depletion.

In the case where there is a finite dielectric mismatch, e.g., $\Delta = 0.95$, as shown in Fig.~\ref{fig:pressure}a, the depletion attraction appears to be more pronounced as more ions are depleted from the slit region due both to ``salt-induced"  image as well as dielectric image repulsions.  The depletion attraction increases with the multivalent salt concentration $c_0$ (compare Figs.~\ref{fig:pressure}a and b), but also with increasing the monovalent salt concentration $n_0$ (Fig.~\ref{fig:pressure}a), which agrees with the depletion effect seen from the density profiles in Fig.~\ref{fig:dens}.  At very close separation,  all salt ions are expelled from the slit and the attractive pressure corresponds to the bulk osmotic pressure. In general one can discern that the implicit simulations follow nicely the explicit simulations, except at very small inter-surface separations, where more explicit details on the screening cloud is needed.

The analytical prediction for the total pressure, Eq.~(\ref{eq:ptot}), is shown in Fig.~\ref{fig:pressure}a (solid curves) along with the explicit and implicit MC data, demonstrating very good agreement with the implicit simulations especially at sufficiently large 1:1 added salt concentration. This confirms the validity of the single-particle approximation underlying the analytical dressed multivalent ion approach for the regime of parameters considered here (i.e., sufficiently large $q$, sufficiently small multivalent ion concentration and sufficiently large monovalent salt concentration). In other words, the many-body effects due to interactions between different (dressed) multivalent ions is negligible and the main contribution results from the self-image interactions as included in the analytical theory above.  The analytical curves show larger disagreement with explicit MC data, especially at small separations, but these deviations amount to less than about 20\% \cite{note_n0}. 

Note that our MC data sometimes reveal a small hump at intermediate separations in Fig.~\ref{fig:pressure}a. This kind of behavior is pronounced at larger valencies and smaller screenings, as shown by our implicit MC simulations in Fig.~\ref{fig:pressure}b. Comparison with standard steric depletion effects~\cite{lekkerkerker} leads us to surmise that a larger valency leads also to a stronger repulsive interaction between dressed multivalent ions that in  turn pushes them out of the bulk and back into the slit, where they act osmotically, thus increasing the repulsive component of the interaction pressure. Note that this  ``soft sphere packing" mechanism is supported by the osmotic pressure measurements on free standing films as well as surface force measurements~\cite{Richetti:92,Bergeron:92,Theodoly:01}, where an oscillatory behavior  is observed between neutral (or charged) flat surfaces in the presence of solutions containing highly charged species (colloidal particles or polyelectrolytes).   

In the final part of this study, we compare the image depletion interactions discussed above with typical values of the vdW contribution that also acts between uncharged surfaces in an additive manner, i.e., all the interactions calculated and analyzed in this work are {\em on top of the vdW interactions} acting across the slit between two dielectric interfaces. In fact the free energy per surface area $S$ of the zero-frequency term of the vdW interactions (which couples to the charge density fluctuations in the salt solution and thus depends on the Debye screening parameter)   is obtained explicitly as~\cite{RudiAliepl}
\begin{equation}
\frac{\beta {\cal F}_0^{\mathrm{vdW}}(a)}{S} =  \int_{0}^{\infty} \frac{Q \,\rmd Q}{4\pi}\,\ln{\left( 1 - \Gamma^2 (Q)\right)},
\label{zerofreq}
\end{equation}
where $\Gamma(Q)$ is defined in Eq.~(\ref{eq:Gamma}). This term corresponds to screened (zero-frequency) electromagnetic field fluctuations, first addressed by Ninham and Parsegian as described in Refs. \cite{Ninham-book,Parsegian}, and later scrutinized at length in, e.g., Refs. \cite{Kjellander1,Kjellander0,Netz2,Rudi-Zeks}.
One should note here that in the above free energy we have already taken into account a cancellation occurring between the standard zero-frequency vdW term and a part of the free energy corresponding to an uncoupled, i.e., uncharged, system, the result of this cancellation being the above formula. 

At small separations, higher-order Matsubara frequencies of the electromagnetic field-fluctuations are expected to 
become important as well, whose contribution to the interaction free energy of the dielectric half-spaces follows from~\cite{Parsegian} 
\begin{eqnarray}
\frac{\beta {\cal F}_{n \geq1}^{\mathrm{vdW}}(a)}{S}\!\! &=&  \!\! {\sum_{n=1}^{\infty}}\int_{0}^{\infty}\!\!\frac{Q \,\rmd Q}{2\pi}\,
   \ln \bigg[ 1+ \Delta^2_{{\mathrm{TM}}}(\imath \xi_n) \,
\rme^{- 4 k(\imath \xi_n) a} \bigg]  \nonumber\\
& & + ~[{\mathrm{TM}} \rightarrow {\mathrm{TE}}],
\label{eq:Free-Energy-Two-sheets}
\end{eqnarray}
with \begin{equation}
k^2(\imath \xi_n) =  Q^2 + \frac{ \epsilon (\imath
\xi_n)  \mu (\imath \xi_n) \xi_n^2 }{c^2},
\label{eq:rho-Matsubara-Two-sheets}
\end{equation}
where $c$ is the speed of light in vacuo, and $\epsilon (\imath \xi_n)$ and $\mu (\imath \xi_n)$ are  the dielectric response function and the magnetic permeability of the slit at imaginary Matsubara frequencies, respectively.   TM and TE correspond to transverse magnetic and transverse electric modes and the  $n$ summation is over  imaginary Matsubara frequencies $ \xi_n=2 \pi n k_{\mathrm{B}} T/\hbar, $ where  $\hbar$ is the Planck constant divided by $2 \pi$. For the sake of simplicity, we assume that for all layers $\mu(\imath \xi_n)=1$ and that the dielectric function for the slit is that for water.  $ \Delta_{{\mathrm{TM}}}(\imath \xi_n)$ and $ \Delta_{{\mathrm{TE}}}(\imath \xi_n)$ are functions of the frequency-dependent response functions of the semi-infinite dielectrics and the water slab whose explicit forms can be found in Ref.~\cite{Parsegian}. The non-zero Matsubara frequency vdW free energy can be written as 
\begin{equation}
\frac{{\cal F}_{n \geq1}^{\mathrm{vdW}}(a)}{S}=-\frac{A_{n\ge 1}}{48\pi a^2}, 
\end{equation}
where $A_{n\ge 1}$ is the corresponding Hamaker coefficient. We take  $A_{n\ge 1}=3$~zJ, which is on the upper bound for the non-zero Matsubara frequency contribution for hydrocarbons interacting across an aqueous medium~\cite{Pod1}. The complete Hamaker coefficient  $A_{n=0} + A_{n\ge 1}$ for, e.g., dimyristoyl phosphatidylcholine (DMPC) and dipalmitoyl phosphatidylcholine (DPPC) multilayers has a typical value in the range of $2.87 - 9.19$~zJ~\cite{hamaker}. 

As seen in Fig.~\ref{fig:pressure}c, the pressure due to the image depletion interactions is of the same order of magnitude as the total vdW (but larger than the zero-frequency vdW contribution) at separations distances about 1~nm and above, especially at large monovalent salt concentrations. One should not forget, however, that the vdW interactions and the effects discussed in this work are additive, hence the total interaction pressure is the sum of the vdW pressure and Eq.~(\ref{eq:ptot}). Obviously the image-related interactions discussed here play as significant a role at the nano scale as the vdW interactions and should thus not be overlooked. 

\section{Conclusion} 

We consider an asymmetric ionic fluid, consisting of multi-species monovalent salt ions and a single species of multivalent ions, confined between two uncharged (neutral) dielectric half-spaces. Specifically we studied the ion distribution and the electrostatic interaction mediated by the ionic fluid between two bounding dielectric interfaces. 

Apart from vdW interactions, the standard DLVO theory of colloidal interactions in an ionic solution \cite{Parsegian,leshouches,Andelman} predicts that electrostatic interactions between neutral dielectrics are non-existent. Any interaction found besides the vdW interactions would thus present a serious challenge to accepted wisdom in the colloidal domain (such electrostatic forces are reported between randomly charged dielectrics that are neutral {\em on the average} but carry a charge disorder component \cite{prl2010,jcp2010,pre2011,epje2012,pre2005}, 
which we do not consider here and focus only on strictly uncharged dielectrics). 

Our purpose was twofold: first, to investigate the nature of the interactions engendered by the dielectric inhomogeneities and salt redistribution between the dielectric interfaces by means of both explicit- and implicit-ion simulations, and to assess  the validity of a simple analytical description that allows us to couple in a consistent way both the ``salt-induced" image effects as well as the dielectric polarization effects. We showed that the ionic mixture of a monovalent salt and a salt with multivalent ions can be decomposed into two parts, which can be treated 
in a single analytical framework. 
We  described the monovalent salt solution on a DH level, which--in other words--means that the monovalent ions are integrated out, leading to a simplified but accurate description for the multivalent ions in the solution. We refer to this 
procedure as ``dressed multivalent ion" approximation. This approximation is valid for highly asymmetric salt mixtures, i.e., when multivalent ions have a large charge valency~\cite{kanduc_dressed1,kanduc_dressed2}. 
Even with the monovalent ion degrees of freedom integrated out, the remaining system containing only the dressed multivalent ions still represents a full many-body problem and additional approximations need to be introduced in order to calculate the partition function of the system analytically. 
To this end, we employed a virial expansion in order to obtain the distribution of multivalent ions and the interaction pressure between the surfaces mediated by them up to the
leading order in multivalent counterion fugacity. This amounts to a simple single-particle theory for dressed multivalent ions which 
 is expected to work for sufficiently small multivalent ion concentration and high enough monovalent ion concentration as we have indeed confirmed by our explicit-ion simulations. 

Although these approaches are based on different rationales, the final outcomes share some qualitative similarities. All the ions are repelled from neutral surfaces due to ``salt-induced" image effects, which are always present for ion-impermeable interfaces, as well as dielectric image effects present only in the case of dielectric inhomogeneities. This leads to ionic depletion in the slit, which results in an attraction between the slabs caused by the osmotic pressure of multivalent ions in the bulk. Our theoretical results are corroborated by explicit and implicit MC simulations, which show
an excellent agreement in the cases where a comparison makes sense.

The magnitude of the effects described in this work is comparable or surpasses the underlying vdW interactions between neutral dielectric interfaces, and should thus perforce be included in colloidal interaction equilibria for uncharged matter in complex multicomponent ionic solutions. 

\section{Acknowledgment} 

We thank R. Kjellander for a critical reading of a previous version of this manuscript and for his many helpful suggestions and comments. 
M.K. is supported by the Alexander von Humboldt Foundation.
A.N. acknowledges support from the Royal Society, the Royal Academy of Engineering, and the British Academy.
J.F. acknowledges support from the Swedish Research Council. 
R.P. acknowledges support from the Agency for Research and Development of Slovenia (ARRS grant P1-0055(C)).


\begin{thebibliography}{99}
\bibitem{note_n0}
It should be noted also that the analytical estimation of the pressure due to the monovalent ions, Eq. (\ref{eq:contact2}), is based on a linear DH approximation and thus, in general, tends to underestimate
the bulk pressure (giving it in the well-known DH form $\beta p_0(a\rightarrow \infty) =n_b-\kappa^3/(24\pi)$) as the bulk salt concentration is increased.
\end{thebibliography}
\end{document}